# Direct Characterization of Band Bending in GaP/Si(001) Heterostructures with Hard X-ray Photoelectron Spectroscopy

Martin Schmid[1]*, Kunie Ishioka[2], Andreas Beyer[3], Benedikt P. Klein[1],

Claudio K. Krug[1], Malte Sachs[1], Hrvoje Petek[4],

Christopher J. Stanton[5], Wolfgang Stolz[3], Kerstin Volz[3],

J. Michael Gottfried[1], Ulrich Höfer[3]

[1] Department of Chemistry and Material Sciences Center, Philipps-University Marburg,
   D- 35032 Marburg, Germany
[2] National Institute for Materials Science, Tsukuba,
   305-0047 Japan
[3] Department of Physics and Material Sciences Center, Philipps-University Marburg,
   D-35032 Marburg, Germany
[4] Department of Physics and Astronomy, University of Pittsburgh, Pittsburgh,
   PA 15260, USA
[5] Department of Physics, University of Florida, Gainesville, FL 32611 USA

*Author to whom correspondence should be addressed

We apply hard X-ray photoelectron spectroscopy (HAXPES) to investigate the electronic structures in ~50-nm thick epitaxial GaP layers grown on Si(001) under different conditions. Depth profiles of the local binding energies for the core levels are obtained by measuring the photoemission spectra at different incident photon energies between 3 and 7 keV and analyzing them with simple numerical models. The obtained depth profiles are in quantitative agreement with the band bending determinations for the same samples in a previous coherent phonon spectroscopic study. Our results demonstrate the applicability of the HAXPES with varying incident photon energy to characterize the electric potential profiles at buried semiconductor heterointerfaces.

## I. Introduction

Band alignment at a semiconductor heterointerface determines functionality and performance of an electronic device. The energy diagram at an abrupt heterointerface between two semiconductors can in principle be estimated from electron affinities of the constituent materials[1]. Actual interfaces, however, are far more complicated due to the formation of new chemical bonds and trap states, intermixing of atoms between the two semiconductors, charge transfer, *etc*., causing the actual band alignment to significantly deviate from simple models. To optimize the performance of a functional interface, it is therefore crucial to determine energy levels and band alignments from experimental measurements.

One method for directly characterizing the electronic energy levels and band structures is photoelectron spectroscopy (PES)[2,3]. Although the binding energy is defined primarily by the chemical state of the atom in a solid from which the photoelectron is emitted, it is also influenced by transient and/or local electronic charges and electric fields[4]. This makes PES an opportune experimental method for the identification of local band structure within semiconductor heterostructures. Conventional PES using UV or soft X-ray radiation (UPS or XPS) probes only the near-surface region, because the energy-dependent inelastic mean free path $\lambda$ of the photoelectrons is in the order of a few nm. By contrast, hard X-ray photoelectron spectroscopy (HAXPES) using incident photon energies between 3 and 10 keV is suitable to probe deeper layers up to several tens of nm, because $\lambda$ increases with $E_k$ in the range of $E_k>100$ eV.[3,5] Using sufficiently high photon energy, one can also tune the depth sensitivity by varying the photoelectron detection angle $\Theta$ with respect to the surface normal. The path of photoelectrons traveling within the material is thereby increased by a factor of $1/\cos\Theta$. Because the scattering probability of photoelectrons is proportional to the path length traversed in the material, tuning the detection angle directly tunes the effective probing depth of the HAXPES measurement. This approach was employed previously by Imura and co-workers to examine band-bending effects in InN epilayers.[6,7] Depending on the experimental convenience such as the small sample sizes, and the available detection geometries and the availability of tunable light sources, one is also able to tune the depth sensitivity by varying the incident photon energy while detecting the photoelectrons at a fixed angle $\Theta$.

## II. Experimental

The heterointerface between GaP and (001)-oriented Si is of particular interest, for fundamental and practical reasons.[8-11]   Because it combines a polar and nonpolar semiconductor with a good epitaxial relationship, the interface has the potential for incorporating ultrafast optical response in Si-based electronic devices.[12]   In addition, both GaP and Si have the conduction band minimum at the X point, enhancing transport across the interface.   Fabrication and characterization of an atomically abrupt and defect-free GaP/Si(001) interface has been a challenge, however.   A previous XPS study on ≤1 nm thick GaP nucleation layers grown on Si(001) observed chemically shifted P and Si subpeaks whose intensity corresponded to one monolayer, suggesting the transition from Si substrate to GaP layer in one monolayer.[11]   A recent TEM study reported for thicker (~50 nm) GaP layers on Si(001) the formation of a GaP-Si intermixed region of ~7 monolayer thickness.[13]   Moreover, antiphase domains (APDs) of GaP arising from the monoatomic steps of the Si substrate are unavoidable if one uses the exact (001)-oriented Si substrate.[14,15]   These features drastically modify the electronic structure across the interface such that simple theoretical models no longer apply.

In our previous study we performed ultrafast coherent phonon spectroscopy to characterize the electronic band profiles across the GaP/Si(001) interfaces for samples that were prepared under different growth conditions.[16]   Because the coherent longitudinal optical (LO) phonons in GaP are photoexcited by transient screening of the surface electric field, the amplitude of the LO phonon-plasmon coupled mode gives a measure of the surface electric field before it is screened by photoexcited carriers.   We obtained the band alignments at the GaP/Si(001) interfaces by modeling the coherent phonon signals with help of theoretical simulations under simple assumptions.   However, without the knowledge of the exact doping level of the GaP films and interfacial defect states, the estimation was still indirect and semi-quantitative.

In the present study, we apply HAXPES to directly probe the electronic band structures in two of the GaP/Si(001) samples examined in our previous study.   We obtain the depth profiles of the binding energies within the ~50-nm thick GaP layers by varying the incident photon energy.   Compared to the variation of the detection angle, this approach proved advantageous in our experiment because of the small sample sizes and the geometrical constraint in the electron detection.   We find that our HAXPES results quantitatively confirm the previously obtained indirect characterization of the electronic potentials by coherent phonon spectroscopy.

The samples studied are nominally undoped GaP layers grown on *n*-type Si(001) substrates under different conditions by metal organic vapor phase epitaxy.[16] The GaP layers were grown at 675°C on a Si(001) substrate with very small (<0.1°) miscut angle (sample I), and at 575°C on a Si(001) substrate with a larger (~2°) miscut angle (sample II).  The film thicknesses are 57 nm for sample I and 45 nm for sample II. The different growth conditions and substrates introduce different densities and shapes of APDs, which were characterized by transmission electron microscopy (TEM). Sample I has self-annihilated (kinked) APDs with heights that are comparable to the film thickness, whereas sample II has only a few APDs whose heights are 5 nm or smaller.[16] From previous coherent phonon spectroscopic measurements combined with theoretical calculations it was concluded that sample I has a steep downward band bending toward the surface, whereas the bands in sample II are nearly flat, as shown in Fig. 1.

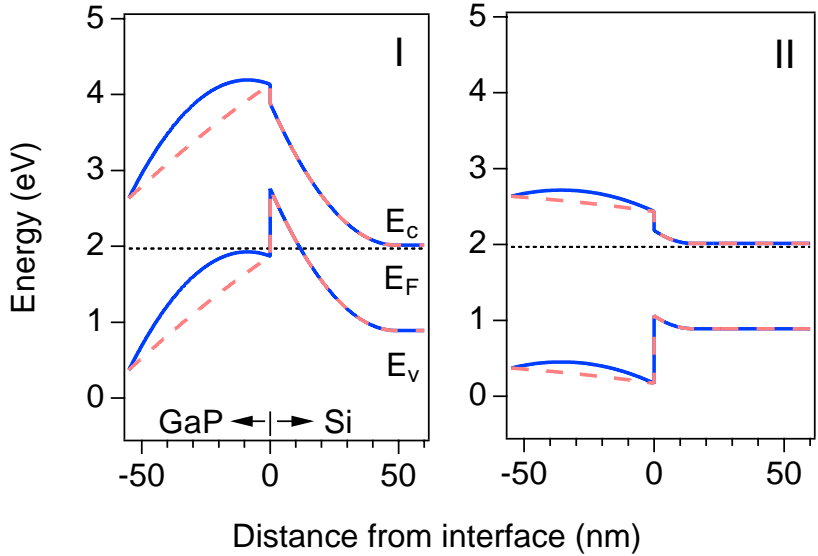

Figure 1.  (Color Online.) Energy band diagrams calculated to reproduce the experimentally obtained Fermi level $E_F$ of GaP/Si samples I and II.  $E_c$ and $E_v$ denote the conduction band minimum and valence band maximum.  Solid and dashed curves assume uniform charge distribution over the GaP layer and localization of charge at the GaP/Si interface, respectively.  Adapted from Ishioka et al.[16].

The HAXPES measurements were performed at the synchrotron radiation facility BESSY-II in Berlin at the High Kinetic Energy (HIKE) photoelectron spectroscopy endstation of the KMC-1 beamline. The beamline is equipped with a double crystal monochromator and is specifically designed for HAXPES experiments.[17,18] Photoelectrons from the sample surface were detected with a VG Scienta R4000 electron analyzer along the surface normal ($\Theta = 0°$). The binding energy $E_b$ of a core level was obtained from the measured kinetic energy $E_k$ of the photoelectrons by the relation $E_b = h\nu - E_k - \phi$, where $\phi$ is the work function of the GaP surface. Incident X-ray photon energies ($h\nu$) between 3 and 7 keV were used to vary the probing depths between 12.9 and 36.5 nm (for the Ga $2p_{3/2}$ signal, see Figures S1 and S2 in the supporting information). The probing depth (or information depth) is defined as three times the inelastic mean free path, $\lambda$, of the photoelectrons (here in GaP). The layers within the probing depth of $3\lambda$ contribute approximately 95% of the PES signal, implying that around 5% of the signal stems from layers beyond the probing depth. To calibrate the HAXPES spectra, we also recorded the Au 4f signal from a gold foil mounted next to the samples for every photon energy. The error in the peak position of the Au 4f peak increases with increasing $h\nu$, from $\pm 0.15$ eV at $h\nu = 3$ keV to $\pm 0.3$ eV at $h\nu = 7$ keV, because the resolution of the X-ray monochromator and the photon flux decrease with increasing photon energy. As a consequence, the peak shapes of the Au4f calibration peaks change such that the peak position can be extracted with limited accuracy only.

### III. Results

Figure 2 compares the P 1s and Ga $2p_{3/2}$ HAXPES spectra of the GaP/Si samples I and II at different photon energies $h\nu$. For sample I, both core level peaks significantly shift towards lower binding energies (downshifts by $0.5 \pm 0.3$ eV for P 1s and $0.64 \pm 0.3$ eV for Ga $2p_{3/2}$) when h$\nu$ increases from 3 to 7 keV, whereas the peak positions of sample II are less dependent on h$\nu$ (downshifts by $0.3 \pm 0.3$ eV for P 1s and $0.15 \pm 0.3$ eV for Ga $2p_{3/2}$) as summarized in Fig. 3a and 3b. Because different photon energies correspond to different probing depths, as we will discuss quantitatively below, the results indicate a sharp electric potential gradient within sample I, but moderate gradient for sample II. These results are congruent with our previous analysis of coherent phonon spectroscopy,[16] which obtained the band profiles plotted in Figure 1. We note that we can in principle expect to find differences in the peak shifts between the P 1s and Ga $2p_{3/2}$ levels, because of the smaller kinetic energy of the P 1s photoelectrons associated with a smaller probing depth for the P 1s level. The random error which is induced in the absolute peak positions by the calibration procedure is, however, larger than the differences in peak shifts between P 1s and Ga $2p_{3/2}$ lines.

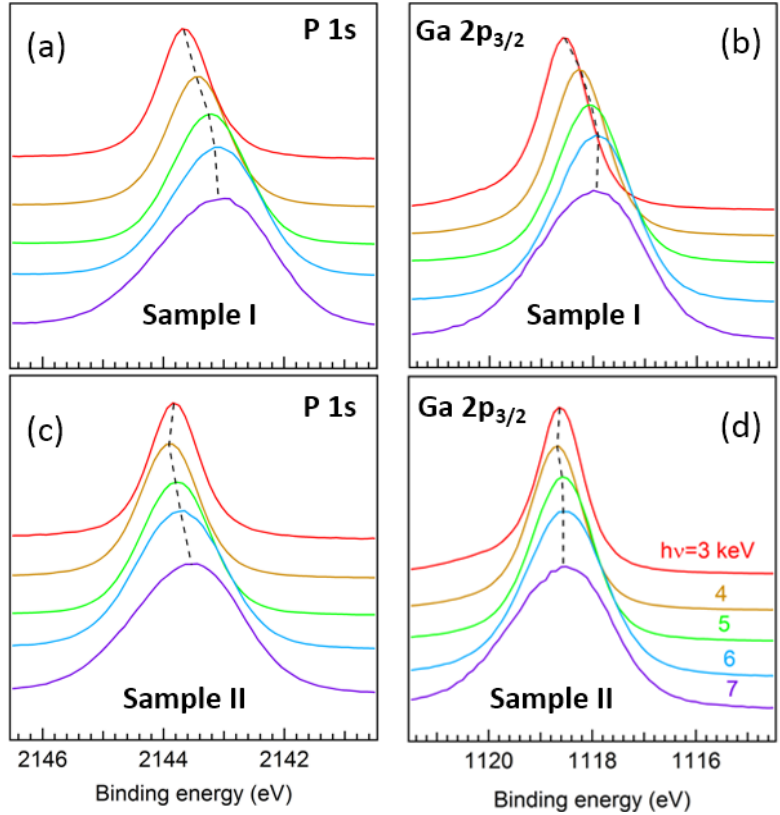

Figure 2. (Color Online.) HAXPES spectra showing the P 1s (a) and Ga $2p_{3/2}$ (b) levels of samples I and II measured with different incident photon energies hv.

The HAXPES spectra contain information from different depths, from the GaP/vacuum surface down to ~3λ at a given photon energy. We estimate the local binding energy $E_b(z)$ as a function of distance from the GaP surface $z$ with the following procedures. We first consider a number of GaP sublayers along the depth direction, each giving a Voigt profile for the HAXPES peak, with 50% gaussian and 50% lorentzian character, at a peak position at $E_b(z)$. The full width $\delta E$ of the Voigt spectral peak is assumed to be dependent on $hv$, from 0.9 eV at $hv$ = 3 keV to 1.67 eV at $hv$ = 7 keV, but independent of $z$ for given $hv$. The photon energy-dependence of the width is based on the experimentally observed broadening of the Au 4f calibration peak in the same photon energy range.  The spectral contribution from each sublayer has a depth-dependent statistical weight of $\exp(-z/\lambda)$ to the depth-integrated HAXPES spectrum. The values for the mean free path $\lambda$ in GaP are estimated to vary linearly according to

$$\lambda[\text{nm}] = 6.23 \text{ nm} + 1.973 \text{ nm/keV} \cdot (h\nu - E_b)[\text{keV}] \qquad (1)$$

which is an extrapolation from $\lambda$ values for kinetic energies below 2 keV to the kinetic energy range in this study (see also Figures S1 and S2 in the supporting information).[19] We note that our estimation of $\lambda$ is supported by the variation of the peak intensity of the Si 1s core level emission from the Si(001) substrate through the GaP film.  The HAXPES spectra from sample II, with the 45-nm thick GaP layer, reveal a small yet visible Si 1s peak from the Si substrate at 1840 eV only when probed at 7 keV and not at lower photon energies (see Figure S3 in the supporting information).  By contrast, the HAXPES spectra from sample I (GaP thickness of 57 nm) show no visible contribution from the Si substrate at any photon energy.  These results are consistent with our estimation that the probing depth for Si 1s photoelectrons ($3 \cdot \lambda_{Si1s}$) is ~49 nm at $h\nu = 7$ keV.[19]

In the following numerical analysis, the peak shift of the (in principal) more bulk sensitive Ga $2p_{3/2}$ signal with increasing photon energy is simulated based on the model described above.   In view of the small number of data points and the fact that the random error on the peak position is larger than the differences between the P 1s and Ga $2p_{3/2}$ lines, the average of the P 1s and Ga $2p_{3/2}$ peak positions, rather than the position of the Ga $2p_{3/2}$ peak alone, is optimized.  This procedure should lower the influence of singluar strong random deviations ('spikes') to the computational results. However, this procedure has only an appreciable effect on the shifts observed for the P 1s and Ga $2p_{3/2}$ lines of sample II at photon energies of 6 and 7 keV.  For all other cases, the P 1s and Ga $2p_{3/2}$ shifts are practically identical and there is nearly no deviation of those values from their average.

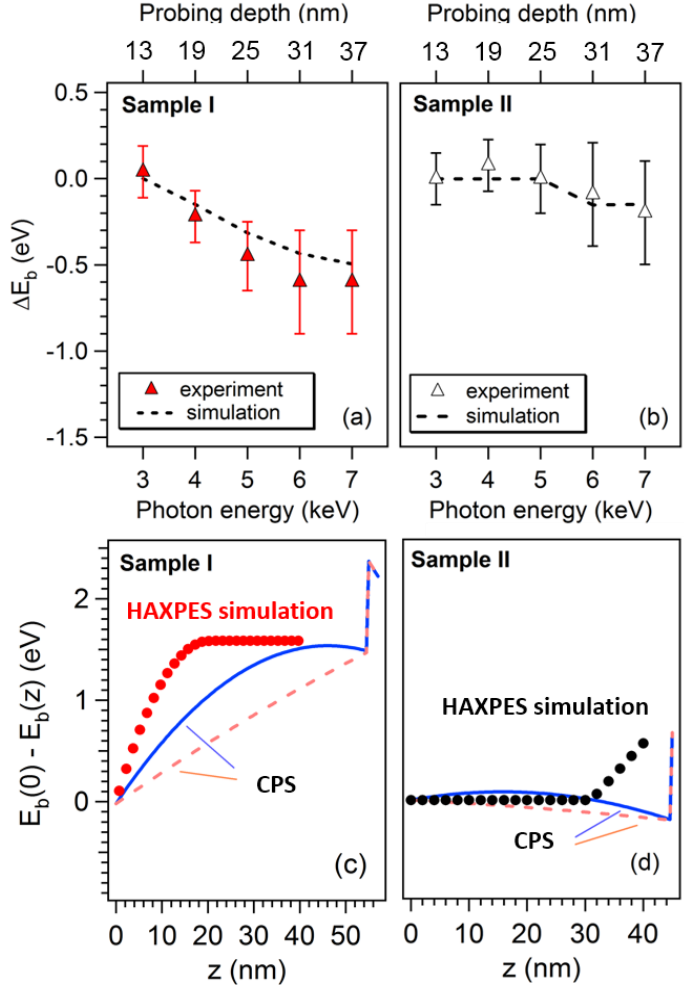

Figure 3. (Color Online.) (a,b) Measured (symbols) and simulated (lines) shifts in the Ga $2p_{3/2}$ binding energy ($\Delta E_b$) with respect to the peak position at $hv = 3$ keV, as a function of the photon energy $hv$; for samples I (a) and II (b). Top axis refers to the probe depth for the Ga $2p_{3/2}$ level. (c,d) Local core level binding energy shift $E_b(0)-E_b(z)$ as a function of distance from surface $z$, obtained from the simulations to reproduce the experimental HAXPES spectra (symbols); for samples I (c) and II (d). Solid and broken curves in (c,d) represent $E_v$ plotted in Figure 1, which are vertically shifted to start from 0 at $z = 0$, CPS stands for coherent phonon spectroscopy.

For sample I, the experimentally observed downshift of the HAXPES peak with increasing $hv$ as shown in Figure 3a indicates a gradual decrease in the local binding energy $E_b(z)$ with increasing $z$. We approximate $E_b(z)$ with a half-parabolic function[20]:

$$E_b(0) - E_b(z) = \Delta\psi_1 - (\Delta\psi_1/d_1^2)(z - d_1)^2 \quad \text{for} \quad 0 \leq z \leq d_1$$
$$E_b(0) - E_b(z) = \Delta\psi_1 \quad \text{for} \quad z > d_1. \quad (2)$$

to model the downward band bending toward the surface,[21] and optimize the parameters $\Delta\psi_1$ and $d_1$ via a genetic algorithm[22,23] to reproduce the peak positions and the widths of the experimentally measured HAXPES spectra.[24] The best fit is obtained with $\Delta\psi_1 = 1.58$ eV and $d_1 = 21$ nm, for which the peak position of the depth-integrated spectra is shown as a function of hv in Figure 3a. The corresponding binding energy variation with $z$, $E_b(0)-E_b(z)$, is plotted in Figure 3c.

The binding energy obtained from our analysis reproduces well the electronic bands calculated in the previous coherent phonon study,[16] as shown in Figure 3c, confirming that the coherent phonon spectroscopy on the valence levels and the HAXPES on the core levels give quantitatively similar estimations on the local electric potential. The previous study calculated two extreme cases, one assuming uniform charge distribution over the GaP layer and the other assuming charges localized in the vicinity of the GaP/Si interface. The results of the present study are consistent with the electric charges being distributed almost uniformly over the whole GaP layer for the nominally undoped sample I. Indeed, their concentration is sufficiently large to cause the surface band bending to saturate before reaching the GaP/Si interface ($d_1 \sim 21$ nm). We can deduce the average charge density $N_1$ using the relation for the surface depletion region for semiconductors with uniform charge profiles:

$$\Delta\psi_1/d_1^2 = e^2 N_1/\varepsilon \quad (3)$$

with $\varepsilon$ being the dielectric constant of GaP ($9.84\times10^{-11}$ As/Vm)[25]. The obtained density, $N_1 \sim 2 \times 10^{18}$ cm$^{-3}$, is in the same range as the value used in the calculation in the case of uniform distribution in the previous study, $9\times10^{17}$ cm$^{-3}$.[16] The origin of the electric charges can be unintentionally doped Si impurities. As shown in Figure S3 in the supporting information, the Si 1s signal consists of two components. The one at 1840 eV, only visible for sample II with the thinner 45 nm GaP layer and at the highest photon energy of 7 keV, stems from the Si substrate. The other component at 1844 eV, which appears at all photon energies, is attributed to a homogeneously distributed Si impurity in the GaP layer. This impurity has a concentration of ~2-3%, according to an analysis of the relative intensities of the Ga, P and Si related signals.

For sample II, the peak positions of the P 1s and Ga $2p_{3/2}$ levels at $h\nu = 3$ keV are very close to those of sample I, indicating that the band offset at the GaP surface with respect to the Si bulk is similar for both samples. In contrast to sample I, however, the average peak position for sample II exhibits almost no shift between $h\nu = 3$ and 5 keV, and only a small downshift at higher photon energies. The observed difference between samples I and II has two important aspects: On one hand, our analysis loses accuracy with increasing $z$, and the observed peak shifts at high photon energies are still within the systematic uncertainty (see Figure 3b). On the other hand, our probing depth extends to the buried GaP/Si interface in case of sample II, which has a thinner GaP layer than sample I. We therefore need to consider the band bending in the vicinity of the GaP/Si interface as well as the GaP surface. For simplicity, we roughly approximate $E_b(z)$ to be constant (i.e., $E_b(0)-E_b(z) = 0$) for $0 \leq z \leq d_2$, and to vary linearly with $z$ according to $E_b(0)-E_b(z) = (\Delta\psi_2/(L-d_2)) (z - d_2)$ for $z > d_2$, with $L$ being the GaP layer thickness. By optimizing the parameters $\Delta\psi_2$ and $d_2$ with the same procedure as for sample I, we obtain the depth-integrated peak position $E_b$ and depth-dependent binding energy $E_b(0)-E_b(z)$ shown in Figure 3b and 3d.

The obtained depth profile suggests a spatially inhomogeneous charge distribution within the GaP layer of sample II. A simple explanation might be given in terms of the excess charges at the APD boundaries consisting of homopolar (Ga-Ga and P-P) bonds, since the APDs in sample II are 5 nm or smaller in height, in contrast to those as high as the GaP layer thickness in sample I. The microscopic structures of the APD boundaries revealed in a previous TEM study[26] are unlikely to sustain sufficient excess charges to affect the electronic band profile, however. Another possible source of the charges is the interdiffusion of Si atoms from the substrate into the GaP overlayer and/or the charged point defects in the vicinity of the GaP/Si interface. While the band structures in Figure 1 were calculated by assuming either uniformly distributed charges (solid curve) or charges nearly localized at the GaP/Si interface (broken curve),[16] an in-between, non-uniform charge distribution could bend the bands strongly only near the GaP/Si interface. The GaP layer thickness (~50 nm) of the present sample does not allow us to more accurately estimate the local, depth dependent potentials associated with band bending near the GaP/Si interface. Based on our results, one would expect that HAXPES with the same incident photon energies could significantly contribute to a quantitative estimation of the interfacial electronic structure for thinner (e.g. 20 nm thick) GaP layers.

## IV. Conclusions

We have investigated the band bending of GaP/Si heterointerfaces by means of HAXPES. The peak shift and the broadening of the P 1s and Ga $2p_{3/2}$ core levels with varying incident photon energy have been analyzed to obtain the depth profile of the local binding energy within the ~50-nm thick GaP layers grown on silicon under two different conditions. For the GaP layer grown on a Si surface with a near-nominal (001) orientation, we have found that the local binding energy varies steeply with depth $z$ and then saturates at $z \sim 30$ nm, which suggests nearly uniform charge distribution at $\sim 2 \times 10^{18}$ cm$^{-3}$. For the GaP layer grown on Si(001) with a larger miscut angle of 2°, by contrast, the local binding energy has been nearly constant up to z ~ 30 nm. The results have quantitatively confirmed the previous indirect estimation of electronic band structure in our coherent phonon spectroscopic study[16]. We have thus demonstrated that HAXPES can be a reliable tool to obtain depth-resolved information on the electronic structures at deeply buried semiconductor heterointerfaces.

## Acknowledgements

This work was supported by the *Deutsche Forschungsgemeinschaft* through SFB 1083, the *Stiftung Stipendien-Fonds des Verbandes der Chemischen Industrie e.V.* (Schmid), as well as by *NSF* Grant DMR-1311845 (Petek) and DMR- 1311849 (Stanton).

## Supporting Information

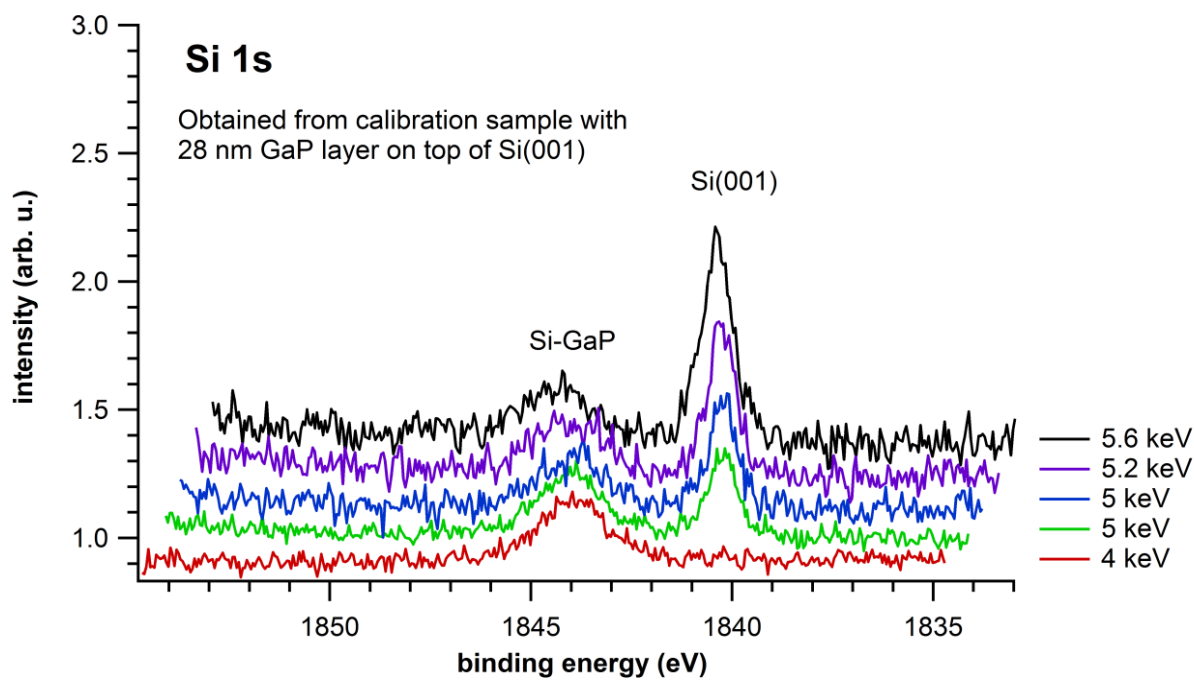

Figure S1: Si 1s region of a calibration sample GaP/Si(001). The peak at 1844 eV emerges from traces of Si in the GaP material while the peak at 1840.2 eV is associated with the Si(001) substrate.

In principle it is possible to determine the exact values for the inelastic mean free paths directly from the observed intensity ratio between Si-GaP and Si(001). Because the concentration of the impurity Si in the GaP material is not *exactly* known (uncertain atomic sensitivity factors at higher energies), such a simple procedure is not possible without a systematic error. However, it is possible to estimate the IMFP at higher kinetic energies from the relative increase in the Si(001) signal if the IMFP at one certain photon energy, e.g. 5keV, is obtained by extrapolation of literature values. The mathematic procedure is as follows:

$$\frac{I_{5\ keV}^{Si(001)}}{I_{X\ keV}^{Si(001)}} = \frac{\exp\left(-\frac{d^{GaP}}{\lambda_{5\ keV}}\right)}{\exp\left(-\frac{d^{GaP}}{\lambda_{X\ keV}}\right)}$$

$$\therefore$$

$$\lambda_{X\ keV} = \left(\frac{1}{\lambda_{5\ keV}} + \frac{1}{d^{GaP}}\ln\left(\frac{I_{5\ keV}^{Si(001)}}{I_{X\ keV}^{Si(001)}}\right)\right)^{-1}$$

Figure S2 shows the result of this extrapolation procedure, where the Si 1s line was chosen as the reference signal for 5 keV – for this spectrum, the IMFP was obtained by extrapolation of literature values.

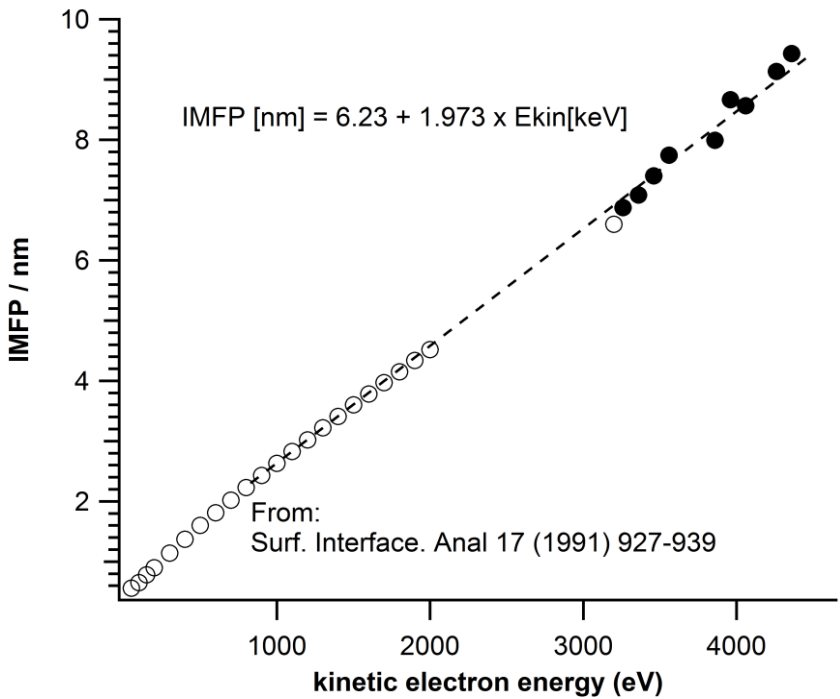

Figure S2: Values for the inelastic mean free path used in this study. The open circles correspond to literature values, the filled circles to our measurements.

In the genetic optimization, the following values for the IMFP have been used:

| $E_{photon}$ | Ekin Ga2p$_{3/2}$ | IMFP Ga2p$_{3/2}$ | Ekin P1s | IMFP P1s |
|---|---|---|---|---|
| 3 keV | 1880 eV | 4.3 nm | 860 eV | 2.29 nm |
| 4 | 2880 | 6.27 | 1860 | 4.26 |
| 5 | 3880 | 8.24 | 2860 | 6.23 |
| 6 | 4880 | 10.22 | 3860 | 8.20 |
| 7 | 5800 | 12.18 | 4860 | 10.17 |

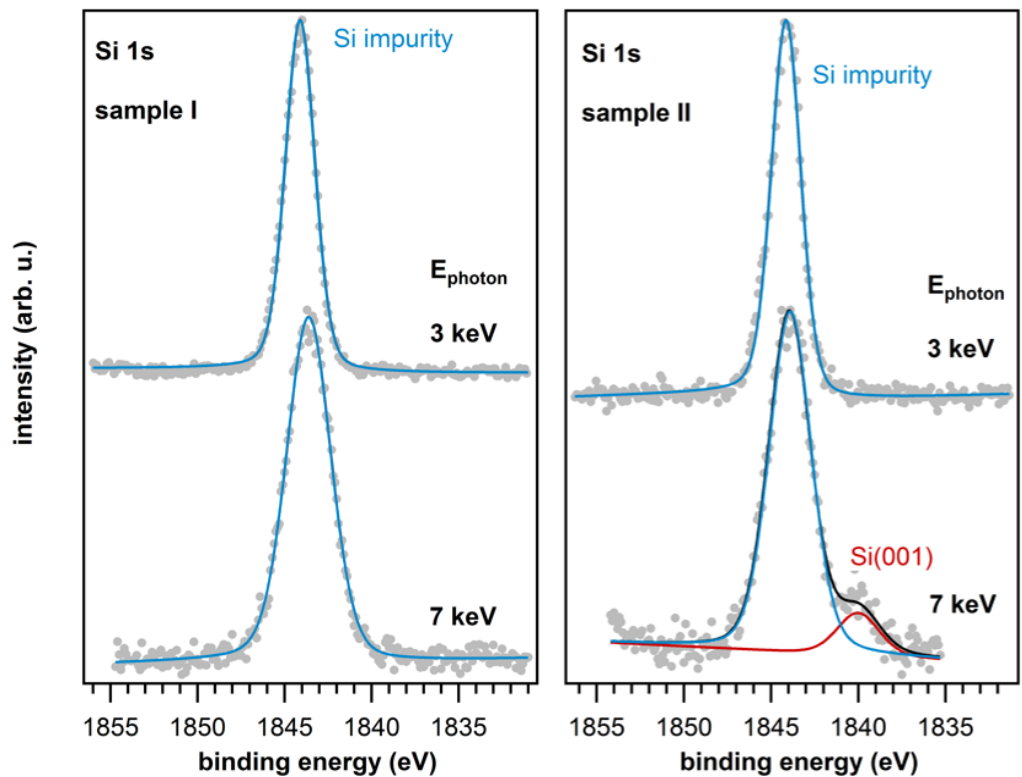

Figure S3: Si 1s region of sample I and II. Sample II has at 7 keV photon energy a small contribution from the Si(001) substrate, whereas the GaP layer in sample I is too thick to allow the detection of any signal from Si(001).